\documentclass[
  reprint,
  amsmath,
  amssymb,
  aps,
  prl,
  letter,
  longbibliography,
]{revtex4-1}

\usepackage[utf8]{inputenc}
\usepackage[T1]{fontenc}
\usepackage[english]{babel}

\usepackage{amsfonts}

\usepackage{booktabs}
\usepackage{csquotes}
\usepackage{url}
\usepackage[pdfusetitle]{hyperref}
\hypersetup{colorlinks=false,pdfborder={0 0 0}}
\usepackage{graphicx}
\usepackage{subcaption}

\usepackage{color}
\newcommand{\rev}[1]{#1}

\newcommand{\dE}{\Delta E}
\newcommand{\SiO}{$\textrm{SiO}_2$}

\newcommand{\Oh}{\textrm{Oh}}
\newcommand{\Ohf}{\textrm{Oh}_\textrm{f}}
\newcommand{\muf}{\mu_\textrm{f}}
\newcommand{\boltz}{k_\textrm{B}}
\newcommand{\EHB}{\Delta G}
\newcommand{\etal}{et al.}
\newcommand{\qO}{q_\textrm{O}}
\newcommand{\qSi}{q_\textrm{Si}}
\newcommand{\ewtheta}{\theta_0^{*}}

\newcommand{\mydeg}[1]{#1^{\circ}}

\newcommand{\powten}[1]{\cdot 10^{#1}}

\renewcommand*{\vec}[1]{\mathbf{#1}}

\begin{document}

\begin{abstract}
We use large-scale molecular dynamics to study dynamics at the three-phase contact line in electrowetting of water and electrolytes on no-slip substrates. Under the applied electrostatic potential the line friction at the contact line is diminished. The effect is consistent for droplets of different sizes as well as for both pure water and electrolyte solution droplets. We analyze the electric field at the contact line to show how it assists ions and dipolar molecules to advance the contact line. Without an electric field, the interaction between a substrate and a liquid has a very short range, mostly affecting the bottom, immobilized layer of liquid molecules which leads to high friction since mobile molecules are not pulled towards the surface. In electrowetting, the electric field attracts charged and polar molecules over a longer range which diminishes the friction.
\end{abstract}

\title{Electrowetting diminishes contact line friction in molecular wetting}
\author{Petter Johansson}
\author{Berk Hess}
\email{hess@kth.se}
\affiliation{Science for Life Laboratory, Department of Applied Physics and Swedish e-Science Research Center, KTH Royal Institute of Technology, Stockholm, Sweden}
\date{\today}
\maketitle

\section{Introduction}
\noindent
Recent developments in the study of liquid droplets spreading on surfaces have shown that the dynamics can be limited by a mixture of inertia, viscous and contact line energy dissipation. The term that dominates the process can be determined from the balance of non-dimensional Ohnesorge numbers, which relate viscous friction to surface tension and inertial forces \cite{DoQuang2015}. These are given by $\Oh{} \equiv \mu / \sqrt{\rho \gamma R}$ and $\Ohf{} \equiv \muf{} / \sqrt{\rho \gamma R}$ where $\mu$ and $\rho$ respectively are the liquid viscosity and density, $\gamma$ the liquid--vapor surface tension, $R$ the initial droplet radius and $\muf{}$ a contact line friction parameter which has units of viscosity. As \citeauthor{DoQuang2015} \cite{DoQuang2015} show \rev{for the initial, rapid wetting phase,} $\Oh{} \gg 1$ correlates with viscous forces dominating the wetting dynamics over surface tension. Similarly, when $\Ohf{} \gg 1$ contact line friction does. This is the case for certain hydrogen bonding \cite{Johansson2015,Johansson2018} or micro-structured substrates \cite{Shiomi2015}.

Surprisingly, applying an electric potential to a droplet, a phenomenon covered under the umbrella of \emph{electrowetting} (see \cite{mugele2005,zhao2013} for comprehensive reviews and \cite{mchale2009,mchale2013,Lomax2016} for some recent experiments), diminishes contact line dissipation \cite{Decamps2000}. Moreover, a recent study using lithographed substrates shows that under electrowetting the wetting can shift from a line friction dominated to a viscously dominated regime \cite{Shiomi2018}. The authors refer to this as an \emph{electrostatic cloaking} of the microscopic substrate features.

It is not yet known how the contact line advancement is affected to reduce the influence of line friction under these conditions. This is not helped by the fact that although models of contact line friction have been proposed for different length scales (including our previous work on molecular wetting \cite{Johansson2018} and of Perrin~\etal{}~on microscopic \cite{Perrin2016}), we lack a holistic understanding of the phenomena.

In this paper we investigate how electrowetting affects contact line friction on a molecular level \rev{in the rapid wetting regime,} using computer molecular dynamics simulations of pure water and an electrolyte solution. We consider how the cloaking effect relates to our previously proposed model of molecular line friction.

\section{Method}
\begin{figure}
  \centering
  \begin{subfigure}[b]{0.45\columnwidth}
    \centering
    \includegraphics[width=\textwidth]{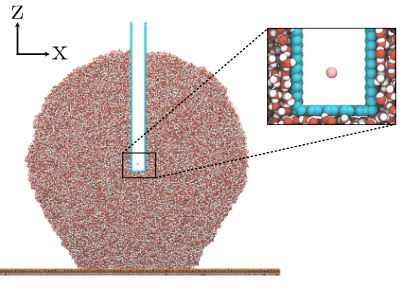}
    \caption{\label{fig:system-2d-sideview}}
  \end{subfigure}
  \begin{subfigure}[b]{0.45\columnwidth}
    \centering
    \includegraphics{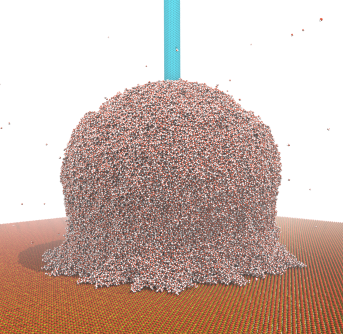}
    \caption{\label{fig:system-3d-wetted}}
  \end{subfigure}
  \caption{\label{fig:system-setup} Electrowetting system with a 15~nm radius water droplet. (a) shows a two-dimensional slice of the system, highlighting the electrode charge. Below the substrate is a layer of counter charges. (b) shows the system during an electrowetting experiment.}
\end{figure}

\noindent
Electrowetting systems were constructed for molecular simulations with three base components: A planar substrate, a liquid droplet and an electrode (figure~\ref{fig:system-setup}).

 The used atomic substrate is a silica-like of \SiO{} electrostatic quadrupoles set in a mono layer with fcc packing.  The equilibrium contact angle $\theta_0$ is set by tuning the atomic charges while keeping the molecules neutral, with $\qSi{} = -2 \qO{}$. We use three sets of charges with equilibrium contact angles of $\mydeg{70}$, $\mydeg{90}$ and $\mydeg{110}$. For the liquid droplet we use both pure water (PW), which hydrogen bonds with the \SiO{} quadrupoles, and a KCl electrolyte solution \cite{Weerasinghe2003,Hess2009} at a concentration of 3M. Since water hydrogen bonds to the silica substrate it is effectively a no-slip substrate \cite{Johansson2015}. The water model is SPC/E \cite{berendsen1987} which has $\rho = 990 \, \mathrm{kg \, m^{-3}}$, $\gamma = 5.8\powten{-2} \, \mathrm{Pa \, m}$ and $\mu = 8.8\powten{-4} \, \mathrm{Pa \, s}$ at the simulated system temperature of 300~K. The initial droplet radii $R_0$ are 7.5 and 15~nm, with respectively around 59,000 and 470,000 liquid molecules. As an electrode a neutral carbon nanotube with radius 1~nm was used, with its interaction parameters tuned to give a $\mydeg{90}$ contact angle.

 An electric potential difference $U$ is created by putting a single fixed charge $Q$ close to the bottom of the electrode and $n$ opposite charges $q_n$ below the planar substrate such that $n q_n = -Q$. These lower charges are free to move in the plane below the substrate. This creates a non-homogeneous electric field directed towards (or away from) the lower end of the electrode. For the 7.5~nm droplets we used $n = 1000$ and $Q = 200e$ while for the 15~nm droplets $n = 4000$ and $Q = \pm 400e$, where $e$ is the electron charge. The larger droplet was run for both a positive and negative potential by switching the charge signs.

\begin{table}
    \centering
    \begin{ruledtabular}
    \begin{tabular}{lccccc}
        Droplet & $R_0$ (nm) & $\theta_0$ (deg) & $\ewtheta{}$ (deg) & $U$ (V) \\
        \hline
        Pure water & 15 & 110 & 65 & $\pm$110 \\
        Pure water & 7.5 & 70 & 56 & 55 \\
        Pure water & 7.5 & 90 & 57 & 55 \\
        Pure water & 7.5 & 110 & 59 & 55\\
        KCl & 7.5 & 90 & 55 & 11 \\
    \end{tabular}
    \end{ruledtabular}
    \caption{\label{tab:final-states} Initial and final droplet states of simulations. $R_0$ is the initial droplet radius and $\theta_0$ the static contact angle for no electric potential. $\ewtheta$ is the static contact angle for the applied potential $U$.}
\end{table}

The charge values were selected to produce a large change in contact angle $\ewtheta$ from the static contact angle $\theta_0$. The electrostatic potential difference $U$ was then measured from the surface to the electrode using the \textsc{pmepot} plugin of \textsc{VMD} \cite{Aksimentiev2005,Humphrey1996}. Contact angles $\theta_0$, $\ewtheta$ and potentials $U$ are reported in table~\ref{tab:final-states}.

Note that experiments of electrowetting display a saturation of the contact angle $\ewtheta$ for increasing potentials $U$. We have not precisely characterized the saturation for our systems, but the experiments appear to be in the saturated regime. A comparison experiment of our 15~nm pure water droplet with $1/2$ the applied potential $U$ does not result in $1/4$ of the force at the contact line as the Young--Lippmann relation \eqref{eq:young-lippmann} predicts, but instead approximately $1/2$. As we will discuss later, the water dipole ordering is high at the contact line which leads to a non-linear dielectric response \cite{Alper1990} that could explain a large part of the saturation. However, analyzing this is outside the scope of this paper.

Simulations were performed using \textsc{Gromacs 2018} \cite{Abraham2015} in double precision with a leap-frog integrator and time step of 2~fs. Short ranged interactions were treated fully up to a cutoff of 0.9~nm. Long ranged electrostatic interactions were treated using the particle-mesh Ewald method which has infinite, periodic interaction range. Periodic boundary conditions were applied along the x and y axes, and repulsive walls were placed at the simulation box edges along z to contain particles to the system. We verified that the periodic boundary treatment does not significantly affect the results by increasing the periodic distances.

Contact angles $\theta(t)$ were measured for each output simulation frame at time $t$ using the approach introduced by Khalkhali~\etal{}~\cite{Khalkhali2017}. The wetting radius $r(t)$ was calculated using a radial density distribution of the bottom layer of water molecules, from its center.

\section{Results and discussion}
\noindent
Wetting simulations were performed in two stages. First the droplets were allowed to relax to their equilibrium states on the substrates while the electrode and below-substrate atoms where uncharged. After equilibration, the charges were increased to their final values over 50~ps using a sigmoid activation function. As the electric field is created, the droplet spreads out to a smaller equilibrium contact angle $\theta_0^{*}$ modeled by the Young--Lippmann relation
\begin{equation}
  \label{eq:young-lippmann}
  \cos{\ewtheta{}} = \cos{\theta_0} - \frac{cU^2}{2 \gamma}
\end{equation}
for the substrate capacitance per unit area $c$. Unlike a prior computational study of electrowetting on gold \cite{yuan2010}, no precursor film is present.

\begin{figure}
  \centering
  \begin{subfigure}[b]{\columnwidth}
    \includegraphics{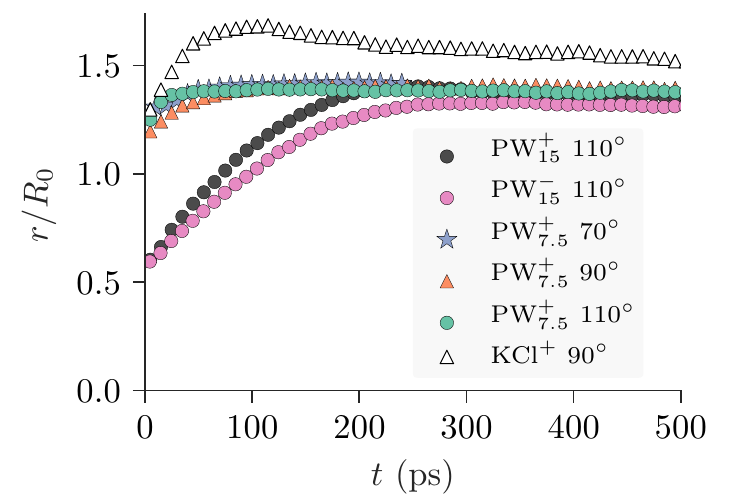}
    \caption{\label{fig:spreading-per-time}}
  \end{subfigure}
  \newline
  \begin{subfigure}[b]{\columnwidth}
    \includegraphics{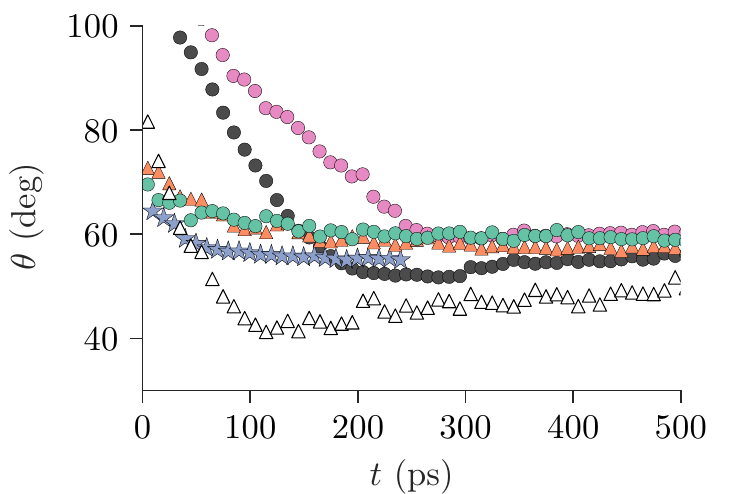}
    \caption{\label{fig:angles-per-time}}
  \end{subfigure}
  \caption{\label{fig:spreading-and-angles} Spreading base radius (a) and contact angle (b) for all systems. The droplet radius is noted for the pure water (PW) systems. The applied potential sign is shown for all systems.}
\end{figure}

\begin{figure}
  \centering
  \includegraphics{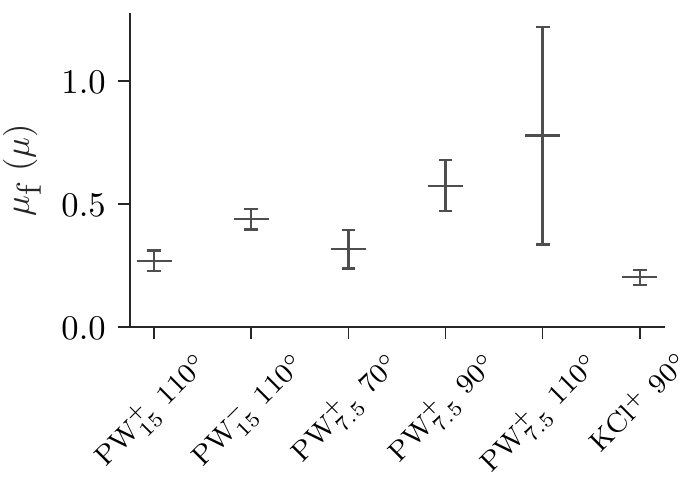}
  \caption{\label{fig:contact-line-friction} Measured mean contact line friction during the rapid spreading phase for all systems. Vertical bars mark $\pm 1$ standard error.}
\end{figure}

We record the base radius $r(t)$ and contact angle $\theta(t)$ starting from the fully applied field. These are presented in Fig.~\ref{fig:spreading-and-angles} with the final states given in table~\ref{tab:final-states}. The final state is reached quickly for all systems, although the 15~nm and electrolyte droplets overshoot and retract to their final $\ewtheta{}$ (not shown in figure). This is particularly noticeable for the KCl system, which has extremely rapid dynamics.
Contact line friction is measured by how much the contact line speed $v$ is damped compared to what we would expect from the Young driving force \cite{Decamps2000,Blake1969,Yue2011}. With $\muf{}$ being the friction parameter the velocity is given by $v = \gamma (\cos{\ewtheta{}} - \cos{\theta}) / \muf{}$. Since we can calculate $v$ from the spreading radius $r(t)$ we estimate the friction parameter for our data sets using this relation. These results are presented in Fig.~\ref{fig:contact-line-friction}, using only \rev{data from the rapid wetting phase. Due to the influence of thermal fluctuations, the 7.5~nm $\mydeg{110}$ droplet gives a relatively high error, but its value is still within the error of the other systems including, notably, both 15~nm $\mydeg{110}$ droplets.} The dynamic evolution of $\muf{}(t)$ and $v(\cos{\theta})$ are both available in the Supplemental Materials \cite{SupplementaryMaterials1,SupplementaryMaterials2}.

\begin{figure*}[t]
  \centering
  \includegraphics{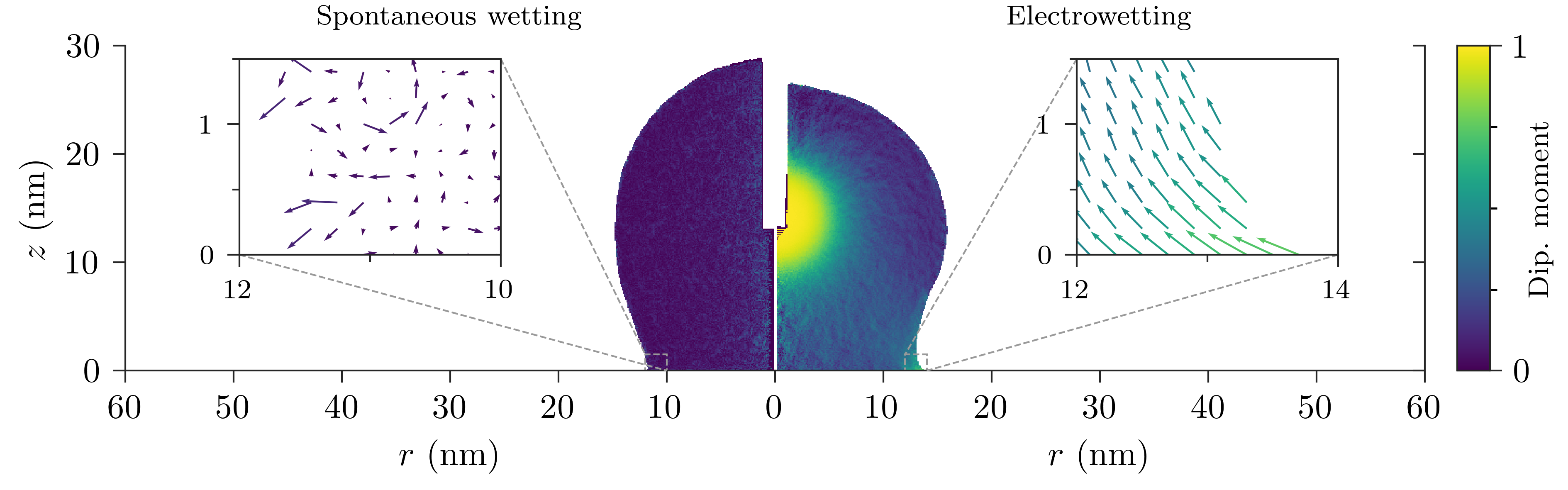}
  \caption{\label{fig:dipole-moments} Radial distribution of measured dipole
    orientation inside droplets during spontaneous (left) and electrowetting
    (right) of pure water. Insets show the averaged dipole direction and
    magnitude at contact lines where the arrows are scaled by a factor of 5 for the spontaneous view. The moment is normalized by the dipole moment of the water molecule.}
\end{figure*}

Consistent with previous studies we see that the line friction is very low for our electrowetting systems and that all systems have very similar amounts of friction, $0.2\text{--}0.8 \mu$. We have previously shown \cite{Johansson2018} that spontaneous wetting of pure water on the $\mydeg{70}$ substrate  without an electric field gives a line friction that increases from $2\text{--}8 \mu$ as the equilibrium is approached, while its average is below $0.5 \mu$ with the electric field. Electrowetting thus decreases the line friction on a molecular level by an order of magnitude, for molecularly flat substrates.

It is interesting to relate our results to the regime map of the Ohnesorge numbers $\Oh{}$ and $\Ohf{}$ \cite{DoQuang2015}, as discussed earlier. For the 15~nm droplet we have $\Oh{} = 0.94$. With a friction factor of $\muf{} = 5 \mu$, as previously measured for spontaneous wetting \cite{Johansson2018}, the same system gives $\Ohf{} = 4.7$. This places the system in the regime where line friction dominates the contact line advancement in spontaneous wetting. With $\muf{} < 0.5$, measured for electrowetting in Fig.~\ref{fig:contact-line-friction}, we shift into the regime where inertial or viscous forces dominate ($\Ohf{} < 1$).

We finally consider where this decrease in line friction originates. Our previous model \cite{Johansson2018} of contact line friction on no-slip substrates proposes that it is an effect of a molecule having to cross an energy barrier $\dE$ of order $\boltz{} T$ in a thermal fluctuation to reach the substrate and advance the contact line, which causes significant friction. This energy barrier stems from the internal hydrogen bonding network between water molecules at the contact line and is given by
\begin{equation}
  \label{eq:energy-barrier}
  \dE = a (\cos \theta + 0.5 \sin \theta)^2 \, ,
\end{equation}
transformed from the reported equation by introducing the Boltzmann factor $\boltz T$ into the exponential (due to it being a thermally activated process) and using some trigonometric identities. At $T = 300 \, \textrm{K}$ the value $a = 1.1 \boltz T$ matched the same silica substrate used here.

That this barrier creates friction comes from another observation: the effective force between the substrate and water has an extremely short range, barely affecting more than the bottom-most layer of water molecules \cite{Johansson2015}. A water molecule cannot feel the attraction from the surface until it gets very close, which means that there is little-to-no assist in crossing the barrier.
In electrowetting this situation changes greatly at the contact line. The potential difference creates an electric field $\vec{E} = -\vec{\nabla} U$. Ions, as in our KCl electrolyte, are directly attracted along this field, leading to a direct assist in crossing the energy barrier.

For neutral molecules, like water, the situation is more complicated. There is no net attraction in a homogeneous electric field but polar molecules with moment $\vec{p}$ experience a force $\vec{F}_p = (\vec{p} \cdot \vec{\nabla}) \vec{E}$ if the field is non-uniform. As water is a dipole this force will be present at the contact line, where the electric field changes abruptly.

\begin{figure}[t]
  \centering
  \includegraphics{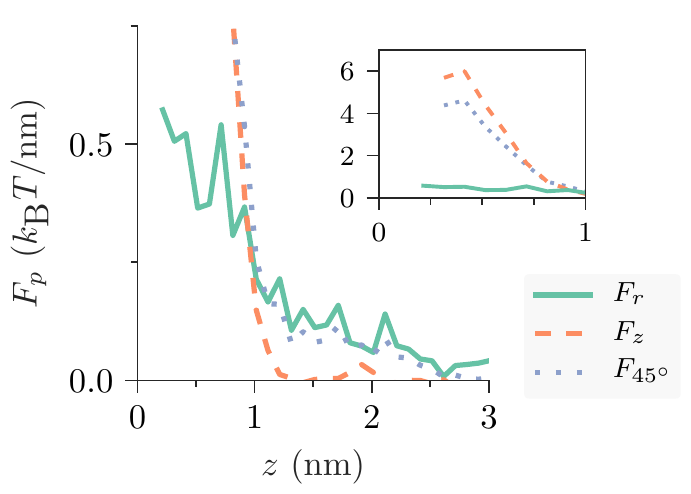}
  \caption{\label{fig:dipole-force} Radial components of the dipole force $\vec{F}_p$ for water molecules at height $z$ above the top substrate atoms. $F_z$ is positive for a force directed \emph{towards} the substrate.}
\end{figure}

Our molecular data allows us to measure both $\vec{E}$ and $\vec{p}$ throughout our system. Figure~\ref{fig:dipole-moments} shows an example radial distribution of the polarization inside a 15~nm pure water droplet during the spreading phase, where the data is averaged over 10~ps. To estimate the influence of $\vec{F}_p$ we calculate it for the same system but after it has reached an equilibrium state with fixed contact angle $\theta_0 = \mydeg{65}$. We calculate it along the droplet interface, averaging over the range inside the radial distribution where the mass is $40\text{--}60\%$ of the bulk value.

Figure~\ref{fig:dipole-force} shows the radial components $F_{r}$ and $F_z$ of $\vec{F}_p$ at different heights $z$ from the top oxygen atom in the \SiO~substrate. We also show a term $F_{\mydeg{45}}$ which is the force projected along a unit vector pointing towards the surface at an angle $\mydeg{45}$. This is to (roughly) represent the force pointing along a path towards the substrate. The z-component of the force is very high close to the surface, owing to the large gradient of the z-component of the electric field in that range. The radial component has a longer tail.

\begin{figure}[t]
  \centering
  \includegraphics{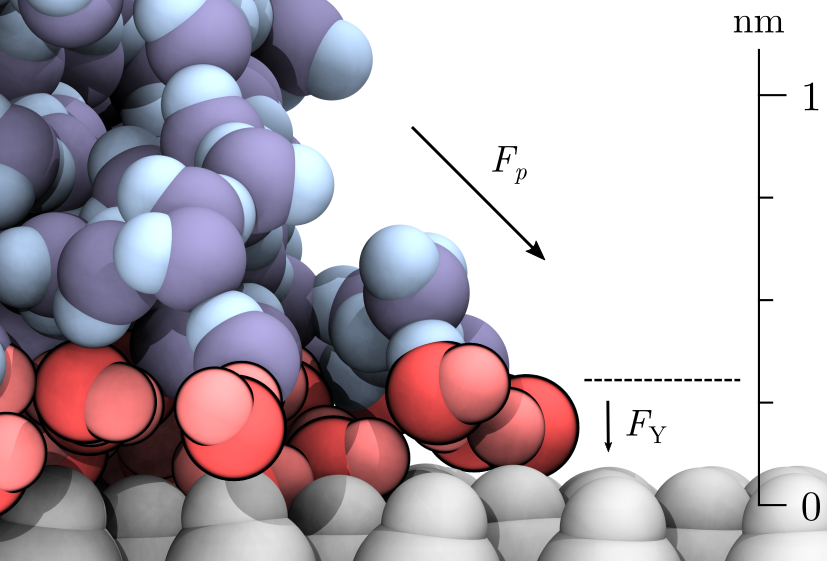}
  \caption{\label{fig:force-range-comparison} The Young force $F_\textrm{Y}$ between the substrate and liquid mostly affects the bottom water layer (shaded red with thick outlines). The dipole force $F_p$ as a longer range, reaching upper molecule layers (shaded blue).}
\end{figure}

How does this compare to the energy barrier $\dE$? With $\theta = \mydeg{65}$ and $a = 1.1 \boltz T$, \eqref{eq:energy-barrier} gives $\dE = 0.84 \boltz T$ and with $\theta = \mydeg{90}$ it is $\dE = 0.28 \boltz T$. Since it is unclear which path a water molecule will move when advancing the contact line we cannot directly calculate its energy gain in the electric field, but even movements of a single water molecule diameter ($\sim 0.25$~nm) with the $F_r$ component yields an estimate of $0.17 \boltz T$, a significant part of the energy barrier at $\mydeg{90}$. Integrating $F_{\mydeg{45}}$ up to 1~nm gives $1.75 \boltz T$. The dipole force term is thus significant over a range of at least 1~nm, which will assist molecules in crossing the energy barrier, if not remove it. We highlight the contrast to the Young force in Fig.~\ref{fig:force-range-comparison}.

Some additional effects related to the polarization at the contact line may contribute to the line friction decrease. As seen in Fig.~\ref{fig:dipole-moments} the water dipoles are highly ordered at the contact line due to the electric field. This ordering may by itself affect the ease of contact line advancement.
We analyze two properties which are affected by the ordering: the hydrogen bonding network of molecules which have to pass the barrier and how bulk viscosity is changed by the high shear stress and ordering.

For the hydrogen bond network we identify the water molecules which are about to advance the contact line by rolling down from the second water layer. The number of hydrogen bonds between these and the surrounding molecules are then counted for the states shown in Fig.~\ref{fig:dipole-moments} by using the \textsc{hbond} tool provided by \textsc{Gromacs} with default settings. For the spontaneous wetting case an average of 2.6 hydrogen bonds were identified per molecule. For the electrowetting case with high dipole ordering we counted 2.0 hydrogen bonds per molecule.

Thus the energy barrier for a molecule rolling to the contact line is lower by up to 0.6 times the free energy of a hydrogen bond $\EHB{}$. For liquid water at room temperature $\EHB{} = 5.7 \, \textrm{kJ} / \textrm{mol} =  2.3 \, \boltz{} T$ \cite{vanDerSpoel2006}, which gives $0.6 \EHB = 1.4 \boltz T$. Note that this number may in part be due to the high ordering and in part due to the transition state having been modified by the electric field and the dipole force $\vec{F}_p$. We can measure the effect but not the cause.

Finally, we consider shear thinning, which may occur for the high shear rates at the contact line during electrowetting. To quantify this effect, we used a simple Couette flow shear setup. We do not observe a significant change in viscosity due to the shear. However, if we additionally apply an electric field which creates 75\% dipole ordering, the viscosity becomes anisotropic. The viscosity with shearing in the direction of the electric field increases by 30\%.
This means that local viscous dissipation will decrease due to the high ordering, but the change is quantitatively much smaller than the weakened hydrogen bonding network described above.

The droplets used here are several order of magnitude smaller than typical droplet sizes. When scaling the droplet size, the voltage can be kept constant to maintain the same contact angle. The electric field drops sharply at the liquid interface, at a molecular length scale \cite{ballenegger2005}. As the reduction in contact line friction is due to the strong electric field and its gradient \emph{at the contact line}, it will be present for larger droplets, consistent with experiments.

We want to note that this purely molecular effect is not the first seen in electrowetting using MD simulations. Daub et al.~have reported on the asymmetry of water molecules yielding dynamics which depend on the sign of the applied potential \cite{daub2007}. Yuan and Zhao on how the precursor film (not present here due to the quadrupole substrate) forms a molecular network with unique transport properties \cite{yuan2010}. And Liu et al.~observed that contact angle saturation occurs as individual molecules are pulled out of the contact line to shield the rest of the interface from the applied potential \cite{liu2012}, an effect that is clearly visible for macroscopic systems. Along with these, our report again highlights how the molecular nature of liquids can influence dynamic and macroscopic behavior.

\section{Conclusions}
\noindent
Through molecular simulation we have verified that application of an electric field decreases contact line friction of smooth, molecular systems, consistent with experiments on Teflon coated and micro-structured substrates \cite{Decamps2000,Shiomi2018}. It is consistent for systems of pure water with different initial radii and an electrolyte. The decrease in line friction is of one order of magnitude, which may shift systems from being dominated by line friction to another dynamic regime.

We have identified two sources at the molecular level for the decrease in friction. The first and main source is that the interaction range between the surface and liquid increases, which largely avoids the high energy barrier that a single layer of water molecules has to cross to advance the contact line. For electrowetting this comes from the electric field and its gradient, which is strong in a region of a few nanometers around the contact line. The other effect is the high ordering of dipoles at the contact line, also due to the strong electric field. This changes the local hydrogen bond network, decreasing the overall number of bonds that water molecules have to break in order to advance the contact line.

Several open questions remain. Most of the above discussion relates to single water molecules independently advancing the contact line. But wetting is a more collective phenomena, where a single advancing molecule pulls along one or more other molecules. Such correlated movements could further influence how we view line friction, but are difficult to study due to the thermal velocity being much higher than the contact line velocity.

\begin{acknowledgments}
We would like to thank Junichiro Shiomi for discussions leading to this work. The research was made possible by funding from the Swedish Research Council (grant no.~2014-4505). Simulations were performed on resources provided by the Swedish National Infrastructure for Computing (SNIC 2018/1-22 and 2019/1-22) at the PDC Center for High Performance Computing. Figures were created using \textsc{Matplotlib} \cite{Hunter2007}.
\end{acknowledgments}

\bibliography{biblio.bib}

\end{document}